\documentclass[twocolumn,secnumarabic,amssymb, nobibnotes, aps, prl]{revtex4-1}

\newcommand{\env}[1]{\texttt{#1}}
\usepackage{graphicx,url,epstopdf,color}

 \definecolor{Black}{named}{Black}
 \definecolor{Blue}{named}{Blue}
 \definecolor{Red}{named}{Red}

\def\la{\mathrel{\mathpalette\fun <}}

\def\fun#1#2{\lower3.6pt\vbox{\baselineskip0pt\lineskip.9pt
  \ialign{$\mathsurround=0pt#1\hfil##\hfil$\crcr#2\crcr\sim\crcr}}}

\begin{document}

\title{Local Magnetic Turbulence and TeV--PeV Cosmic Ray Anisotropies}

\author{Gwenael~Giacinti$^{1,2,3}$}
\author{G\"unter~Sigl$^1$}
\affiliation{$^1$ II. Institut f\"ur Theoretische Physik, Universit\"at Hamburg, Germany}
\affiliation{$^2$ Institutt for fysikk, NTNU, Trondheim, Norway}
\affiliation{$^3$ AstroParticle and Cosmology (APC, Paris), France}

\begin{abstract}
In the energy range from $\sim10^{12}\,$eV to $\sim10^{15}\,$eV, the Galactic cosmic ray flux has anisotropies both on large scales, with an amplitude of the order of 0.1\%, and on scales between $\simeq10^\circ$ and $\simeq30^\circ$, with amplitudes smaller by a factor of a few. With a diffusion coefficient inferred from Galactic cosmic ray chemical abundances, the diffusion approximation predicts a dipolar anisotropy of comparable size, but does not explain the smaller scale anisotropies. We demonstrate here that energy dependent smaller scale anisotropies naturally arise from the local concrete realization of the turbulent magnetic field within the cosmic ray scattering length. We show how such anisotropies could be calculated if the magnetic field structure within a few tens of parsecs from Earth were known.
\end{abstract}


\maketitle
%\tableofcontents

\vspace{3pc}

% Comment if separate title page not required
%\maketitle

\textit{Introduction.}---In the TeV--PeV energy range, statistically significant anisotropies in the distribution of Galactic Cosmic Ray (CR) arrival directions on the sky have been reported both on large and small scales by several experiments, such as Super-Kamiokande~\cite{Guillian:2005wp}, Tibet-III~\cite{Amenomori:2006bx}, Milagro~\cite{Abdo:2008aw,Abdo:2008kr}, ARGO-YBJ~\cite{Vernetto:2009xm} and IceCube~\cite{Abbasi:2010mf,Toscano:2011dc}. On the largest scales these anisotropies have an amplitude of the order of $\sim0.1$\%, and are smaller by a factor of a few to ten on scales between $\simeq10^\circ$ and $\simeq30^\circ$. They are detected up to $E \simeq400$\,TeV by IceCube~\cite{Abbasi:2011zk} and EAS-TOP~\cite{Aglietta:2009mu}. While they appear strongly energy-dependent~\cite{Toscano:2011dc}, with Milagro even finding a localized excess with a different spectrum~\cite{Abdo:2008kr} than for the rest of the sky, they seem to be stable in time~\cite{Amenomori:2010yr}.

The large scale anisotropy can be explained within the diffusion approximation. The transport of charged CRs in the Galactic magnetic field is diffusive at least up to $E \simeq 10^{16-17}\,$eV. The gyroradius of particles of momentum $p$ and charge $Ze$ in a magnetic field of strength $B$,
\begin{equation}
r_g(p)\simeq\frac{p}{eZB}\simeq1\,\left(\frac{p/Z}{10^{15}\,{\rm
      eV}}\right)\,\left(\frac{B}{\mu{\rm G}}\right)^{-1}\,{\rm pc}\,,
\label{eq:gyro2}
\end{equation}
is indeed smaller than the largest length scales, $L_{\max}=100-300$\,pc, on which the turbulent component of the Galactic magnetic field is believed to vary. CRs predominantly scatter on those turbulent field modes whose wave vectors ${\bf k}$ satisfy $|{\bf k}| \sim 2 \pi /r_g(p)$. As a result, the CR propagation in the Galaxy resembles a random walk. It can be considered as a Markovian process in which CRs ``lose'' memory of their past trajectory on distance scales of the order of the scattering length on magnetic field inhomogeneities $\lambda(p)$, which here is essentially the length scale beyond which the CR propagation direction becomes uncorrelated with its original direction.

In the diffusion approximation, an inhomogeneous density $n({\bf r},p)$ of CRs with momenta $p$ leads to a dipole vector given by~\cite{bbdgp,Candia:2003dk}
\begin{equation}
  \hbox{\boldmath$\delta$}(p)\simeq - \frac{3}{c_0}\frac{{\bf j}}{n}=
  \frac{3D(p)}{c_0}\frac{\hbox{\boldmath$\nabla$} n}{n}\,,
\label{eq:diff_dipole}
\end{equation}
where ${\bf j}({\bf r},p)=-D(p)\hbox{\boldmath$\nabla$} n$ is the CR current corresponding to the diffusion coefficient $D(p)=\lambda(p)/3$, here assumed to be homogeneous, and $c_0$ is the speed of light. A nearby source would lead to a dipole oriented towards the source. The contribution of several recent nearby sources to the CR flux at Earth would be a superposition of dipoles and thus again a dipole.
%The observed dipolar amplitude of $\simeq 0.1$\% can be explained within the diffusion approximation:
Eq.~(\ref{eq:diff_dipole}) yields the estimate $|\hbox{\boldmath$\delta$}(p)|\sim D(p)/R$ where $R$ is of the order of the distance to the closest sources. $D(p)$ can be inferred from the chemical abundance ratios of secondary to primary nuclei, such as boron to carbon, and from the anti-proton fraction~\cite{Strong:1998pw}. This gives fits of the form~\cite{DiBernardo:2009ku,Blasi:2011fi}
\begin{equation}\label{eq:diff_coeff}
D(p)\simeq10^{28}\left(\frac{p/Z}{3\,{\rm GeV}}\right)^\delta\left(\frac{z_0}{\rm kpc}\right)
\,{\rm cm}^2\,{\rm s}^{-1}\,,
\end{equation}
where $z_0\sim 1\,$kpc is the scale height of the Galactic disk, and $\delta\simeq0.45$. Theoretically, $\delta=2-\alpha$, for a turbulent field with spectral index $\alpha$: $\delta=1/3$ for a Kolmogorov spectrum. With Eq.~(\ref{eq:diff_dipole}) and $R\simeq z_0$ this yields
\begin{equation}\label{eq:dipole_num}
|\hbox{\boldmath$\delta$}(p)|\sim\mbox{few}\times10^{-3}\,\left(
   \frac{p/Z}{20\,{\rm TeV}}\right)^\delta\,.
\end{equation}
More detailed treatments such as those in Refs.~\cite{Erlykin:2006ri,Blasi:2011fm} give similar values and have found that such large scale anisotropies contain the signature of the few most nearby and recent sources. For a given source distribution, $|\hbox{\boldmath$\delta$}(p)|$ may strongly fluctuate around its average scaling with $p^{\delta}$~\cite{Blasi:2011fm}. This can explain why the anisotropies measured by IceCube~\cite{Toscano:2011dc} are smaller at 400\,TeV than at 20\,TeV. Other works have proposed that the large scale anisotropies may be due to a combined effect of the regular and turbulent Galactic Magnetic Field (GMF)~\cite{Battaner:2009zf}, or to local uni- and bi-directional inflows~\cite{ICRC361}.

However, the intermediate and small scale anisotropies remain hard to explain. The nearest objects believed to contribute to Galactic CRs (such as Vela, Geminga or the Gum nebula) are at distances $\gtrsim 200\,$pc, and thus many gyroradii away, see Eq.~(\ref{eq:gyro2}). Therefore, anisotropies at scales smaller than the dipole should not correlate with the directions towards possible sources. This would only be the case in the presence of additional effects~\cite{Salvati:2008dx,Drury:2008ns}: For instance, Ref.~\cite{Drury:2008ns} supposed the existence of large scale structures in the GMF aligned with the sources, leading to magnetic funneling of CRs. Anisotropic MHD turbulence in the interstellar magnetic field has also been proposed as a candidate to create such small scale anisotropies in the CR arrival directions at Earth~\cite{Malkov:2010yq}. On the contrary, Ref.~\cite{Lazarian:2010sq} assumed that CRs may be accelerated very locally from magnetic reconnection in the magnetotail.

\textit{Small scale anisotropies from local magnetic turbulence.}---In this Letter we show that energy-dependent medium and small scale anisotropies necessarily appear on the sky, provided a large scale anisotropy exists, either dipole --for instance from the inhomogeneous source distribution-- or dipole and quadrupole, such as in IceCube data~\cite{Toscano:2011dc}. The small scale anisotropies are due to the structure of the local turbulent GMF, typically within the scattering length $\lambda(p)$ from Earth. Since the turbulent field modes relevant for cosmic ray scattering depend on the CR rigidity and, therefore, the local volume of the relevant turbulent field is energy-dependent, the small scale anisotropies must be energy-dependent.
%The main advantage of our theoretical suggestion is that it is generic:
Any --isotropic-- turbulent magnetic field would generically produce small scale anisotropies. The diffusion approximation cannot describe them because it averages over different turbulent magnetic field realizations. It thus averages out small scale anisotropies that are uniquely created by the concrete local realization of the field. Neither anisotropic turbulence, nor directional correlations with source directions on the sky, are required to explain the data.
%Yet if such effects are at work, they would appear on top of the effect described here.

CR trajectories in three-dimensional turbulent fields can be regarded as random walks. For times $\gg\lambda(p)/c_0$ the CR transport can be considered as a Markovian process where CRs diffuse on random scattering centers. Let us consider a \textit{point-like source} emitting particles in all directions. For distances from the source $\la \lambda(p)$, the propagation in the local turbulent magnetic field still has memory: The particle trajectories are locally determined by their initial directions, and a very small change of the initial angle would not lead to very significantly different trajectories for distances $\la \lambda(p)$.
%Within this small volume, some trajectories would tend to cross on one side of spheres of radii $\sim \lambda$, while some others would cross on the other side.
For distances to the source $\gtrsim \lambda(p)$, the process becomes Markovian: CRs on nearby trajectories virtually lose memory of times longer ago than $\sim \lambda(p)/c_0$, and their propagation is uncorrelated with initial conditions.

Since Galactic CR energy loss can be neglected and CRs can be back-tracked, the situation is the same for a \textit{point-like observer} of size $\ll \lambda(p)$, such as Earth. Therefore, the small scale anisotropies that are observed at Earth are due to the propagation within the very local interstellar turbulent magnetic field, within the ``sphere'' of radius $\sim \lambda(p)$. The hot spots at Earth are regions where particles are statistically more connected to parts of the ``sphere'' where the dipolar flux impinging from outside the sphere is larger, and cold spots are connected with the part of the external dipole which has a deficit. We also note that the motion of the Sun in the Galaxy cannot smooth out the anisotropies because the distance traveled during the experiment lifetime is $\ll\lambda({\rm TeV})$. Moreover, the local turbulence configuration cannot change noticeably on time scales of $\sim \lambda(p)/c_0$.

\begin{figure}
\begin{center}
\includegraphics[width=0.235\textwidth]{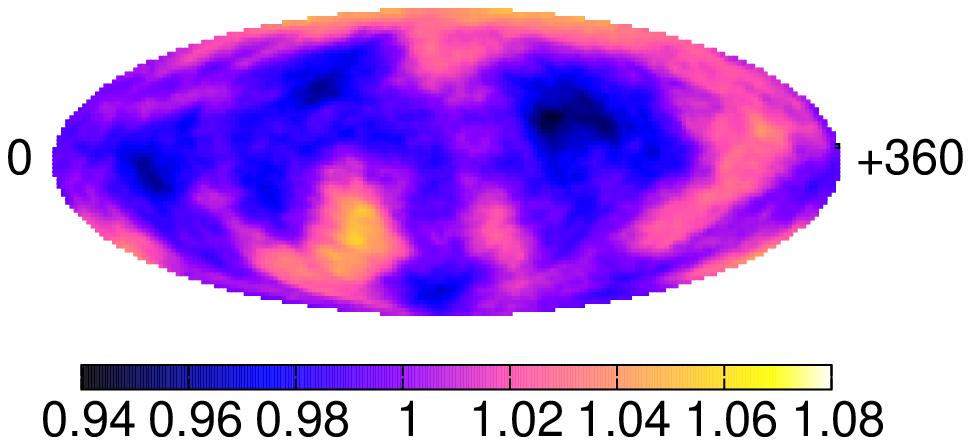}
\includegraphics[width=0.235\textwidth]{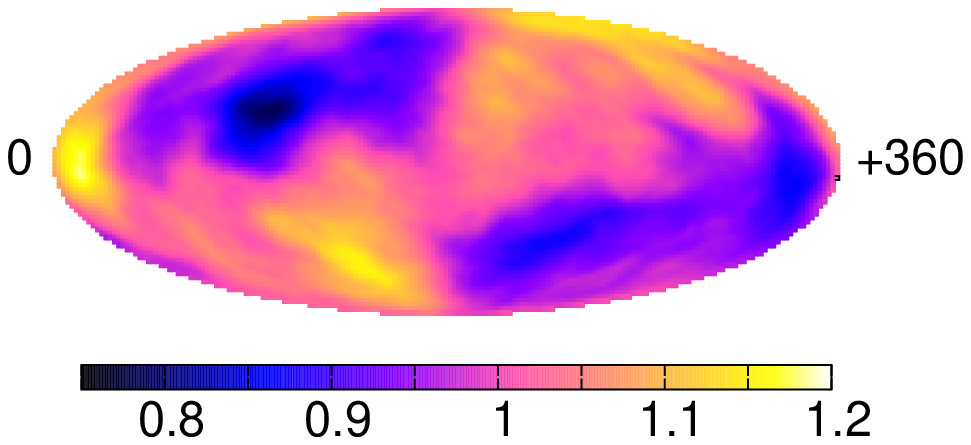}
\end{center}
\caption{Renormalized CR flux predicted at Earth for a concrete realization of the turbulent magnetic field, \textit{after subtracting the dipole}~and smoothing on $20^{\circ}$ radius circles. Primaries with rigidities $p/Z=10^{16}$\,eV (\textit{left panel}) and $5 \times 10^{16}$\,eV (\textit{right panel}). See text for the field parameters and boundary conditions on the sphere of radius $R=250$\,pc.}
\label{AllSkySmallScales}
\end{figure}

To put this more quantitatively, let us imagine two spheres around Earth with radii $r\sim \lambda(p)$ and $R\gg \lambda(p)$, respectively. Back-tracking trajectories of total momentum $p$ from Earth in a given direction parametrized by the unit vector ${\bf n}$ will give a unique point on the sphere of radius $r$, represented by the unit vector ${\bf G}_p({\bf n})$. The function ${\bf G}_p$ is determined by the magnetic field within the small sphere and depends on $p$. It is smooth because trajectories are still non-Markovian on scales $\lesssim \lambda(p)$. When back-tracking further, memory of initial conditions is largely lost, in particular when ${\bf n}$ is averaged over the experimental angular resolution. Still, the expectation value of the crossing point $R{\bf n}_R$ after diffusion within the large sphere must be $\left\langle R{\bf n}_R\right\rangle\simeq\lambda(p){\bf G}_p({\bf n})$. Therefore, the probability $\rho_p({\bf n},{\bf n}_R)$ for a particle emitted in direction ${\bf n}$ at Earth and back-tracked to cross the large sphere at $R{\bf n}_R$ can be estimated as
\begin{equation}\label{eq:rho}
  \rho_p({\bf n},{\bf n}_R)\simeq\frac{1}{4\pi}\left[1+{\bf G}_p({\bf n})\cdot{\bf n}_R\frac{3\lambda(p)}{R}\right]\,,
\end{equation}
which is normalized, $\int\rho_p({\bf n},{\bf n}_R)d^2{\bf n}_R=1$, when integrated over the sphere. If $F_R({\bf n}_R)$ is the differential intensity impinging from outside the large sphere at point $R{\bf n}_R$, the flux $F({\bf n})$ seen at Earth in direction ${\bf n}$ will then be given by
\begin{equation}\label{eq:F_earth}
   F({\bf n})=\int F_R({\bf n}_R) \rho_p({\bf n},{\bf n}_R)d^2{\bf n}_R\,.
\end{equation}
We here assume a uniform gradient of CR density such that $F_R$ is the sum of an isotropic piece $F_0$ and a dipole of relative strength $\Delta$ in $z-$direction ${\bf e}_z$,
\begin{equation}\label{eq:F_i}
   F_R({\bf n}_R)=F_0\left[1+\Delta\,({\bf n}_R\cdot{\bf e}_z)\right]\,.
\end{equation}
This corresponds to $\hbox{\boldmath$\nabla$} n/n=(\Delta/R)\,{\bf e}_z$ and, according to Eq.~(\ref{eq:diff_dipole}), results in a dipole $\hbox{\boldmath$\delta$}(p)\simeq\lambda(p)(\Delta/R)\,{\bf e}_z$ observed at Earth. At the same time, the flux at Earth, Eq.~(\ref{eq:F_earth}), contains multipoles $a_{lm}\equiv \int F({\bf n})\,Y_{lm}({\bf n})d^2{\bf n}/F_0$, normalized to $F_0$, given by
\begin{equation}\label{eq:a_lm}
  a_{lm}(p)=|\hbox{\boldmath$\delta$(p)}| \int {\bf G}_p({\bf n})\cdot{\bf e}_z\,Y_{lm}({\bf n})d^2{\bf n}\,,
\end{equation}
where $Y_{lm}({\bf n})$ are the general spherical harmonic functions. Eq.~(\ref{eq:a_lm}) follows after performing the integration over ${\bf n}_R$ and expressing $\Delta$ in terms of the dipole observed at Earth. In the absence of deflection within a distance $\lesssim \lambda(p)$ from Earth, ${\bf G}_p({\bf n})={\bf n}$ and with $Y_{10}({\bf n})=\left[3/(4\pi)\right]^{1/2}{\bf n}\cdot{\bf e}_z$, Eq.~(\ref{eq:a_lm}) gives $a_{lm}(p)=\delta_{l1}\delta_{m0} |\hbox{\boldmath$\delta$(p)}|\left[(4\pi)/3\right]^{1/2}$. This yields $F({\bf n})=|\hbox{\boldmath$\delta$(p)}|F_0{\bf n}\cdot{\bf e}_z$, corresponding to diffusion with only the dipole present. As in a gas, particles may need several scatterings before losing memory of their momentum. Since the positions of scattering centers in gases are time-dependent, the time-averaged behavior as seen by a given observer is equivalent to particles traveling on straight lines of length $\sim\lambda(p)$ (\textit{i.e.}~${\bf G}_p({\bf n})={\bf n}$) and losing memory after each scattering, also known as ``molecular chaos''. In the case of a concrete local turbulent field realization within $\sim\lambda(p)$, the function ${\bf G}_p({\bf n})\neq{\bf n}$ in general has structures on all angular scales. Eq.~(\ref{eq:a_lm}) predicts $|a_{lm}(p)|\sim |\hbox{\boldmath$\delta$}(p)|$: Clearly, higher multipoles should be of the same order as the dipole impinging from outside the sphere around Earth. As shown in Fig.~\ref{AllSkySmallScales}, the smaller scale anisotropies are much more rigidity-dependent than the dipole. Since the CR composition is not pure proton, and since experiments present data with ``broad'' energy distributions and uncertainties $\Delta p/p \sim 0.3$, small scale anisotropies sum up in a non-constructive way, which typically leads to smaller amplitudes than that of the dipole by a factor of a few to ten. If the structure of the magnetic field and thus the function ${\bf G}_p$ were known within a distance $\sim\lambda(p)$ around Earth, Eq.~(\ref{eq:a_lm}) would even allow to make concrete predictions for the small scale anisotropies. It is also clear that the predicted anisotropies should not depend significantly on $R\gtrsim\lambda(p)$ and should thus converge quickly with increasing $R$. Below we confirm this with numerical simulations. In practice, the Alfv\'enic part of the turbulence is anisotropic~\cite{Goldreich:1994zz} and scattering is mostly dominated by fast modes~\cite{Yan:2002qm}. The condition ${\bf G}_p({\bf n})\neq{\bf n}$ always holds, which systematically ensures the appearance of small scale anisotropies even if scattering is not isotropic. If the diffusion tensor averages to isotropic diffusion on scales $\gg \lambda(p)$, Eqs.~(\ref{eq:rho}) \textit{ff.} are unchanged.

\textit{Numerical simulations.}---We back-track CRs in concrete turbulent magnetic field configurations. The turbulent magnetic field is generated on three-dimensional grids with the method presented in Refs.~\cite{DeMarco:2007eh,Giacinti:2011uj,Giacinti:2011ww}. We use a Kolmogorov spectrum between $L_{\min} \ll r_g$ and $L_{\max}=150$\,pc and with an rms strength $B_{\rm rms}=4\,\mu$G. CRs are injected with isotropically distributed random directions from Earth, located in the Galactic plane $XY$ at $(0,8.5\,{\rm kpc})$, until they cross a sphere of radius $R=250\,{\rm pc}>L_{\max}$ around Earth. Trajectories are considered as flux tubes and a CR crossing the sphere at position $R{\bf n}_R$ is weighted by Eq.~(\ref{eq:F_i}), except that here the dipole is taken towards ${\bf e}_Y$. We take $|\hbox{\boldmath$\nabla$} n/n|=1/(290\,{\rm pc})$ and $p/Z=10,\,50$\,PeV, so as to detect the predicted effect with reasonable statistics.

For this set of parameters we find the expected dipole strength $|\hbox{\boldmath$\delta$}(p)|\sim 6$\% and direction at Earth. Its exact direction and amplitude slightly depend on the concrete realization of the local turbulent field and on the position of the observer, for the same reason as the appearance of small scale anisotropies. Since the dipole scales as $|\hbox{\boldmath$\delta$}(p)|\propto|\hbox{\boldmath$\nabla$} n/n|p^{1/3}$ for Kolmogorov turbulence, taking a more realistic gradient $|\hbox{\boldmath$\nabla$} n/n| \sim 1/(1\,{\rm kpc})$ and energy $p \sim 10\,$TeV would rescale its amplitude to $\sim 0.1$\%. We subtract the dipole from the predicted flux at Earth and show the relative residual intensity maps in Fig.~\ref{AllSkySmallScales}
%(left panel: 10\,PeV; right panel: 50\,PeV),
after smoothing on $20^{\circ}$ radius circles on the sky. The statistical fluctuations in these computations are below $\simeq 0.5$\%. One can see statistically significant features of various amplitudes and shapes, some of which may well resemble the data. Their significance trivially depends on the smoothing radius. Varying the magnetic field realization we find that, while the dipole only slightly varies, the shapes and positions on the sky of the small scale features are strongly realization dependent. In the left panel of Fig.~\ref{AllSkySmallScales} there is a ``hot spot'' with twisted shape and amplitude $\simeq 6$\% in the lower left quadrant and a ``cold spot'' ($\simeq -6$\%) in the upper right quadrant. As predicted, the amplitude is comparable to that of the dipole. At five times larger rigidity the features are strongly different, as seen in the right panel of Fig.~\ref{AllSkySmallScales}. For instance, the cold spot at 10\,PeV in the upper right quadrant is transformed into a hot spot. The energy-dependence of features may also account for the different spectrum~\cite{Abdo:2008kr} seen in the Milagro hot spot. The amplitudes of fluctuations at 50\,PeV are larger than at 10\,PeV here, because we kept the same gradient.

\begin{figure*}[!t]
  \centerline{\includegraphics[width=0.235\textwidth]{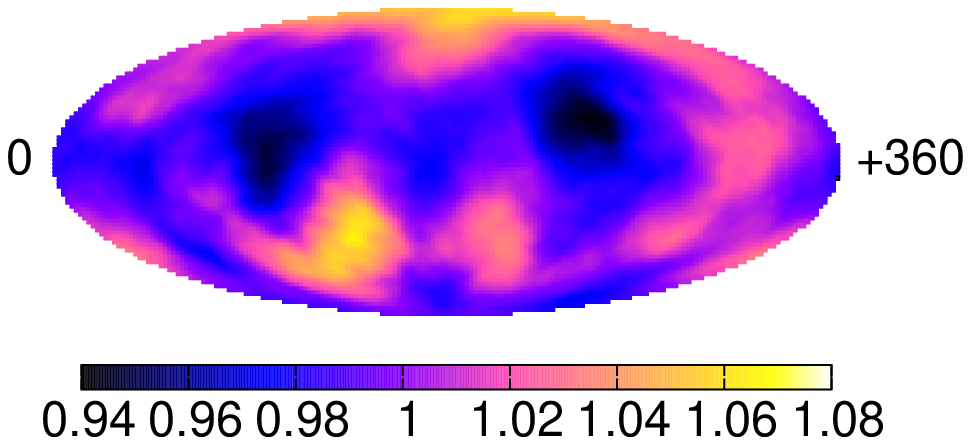}
              \hfil
              \includegraphics[width=0.235\textwidth]{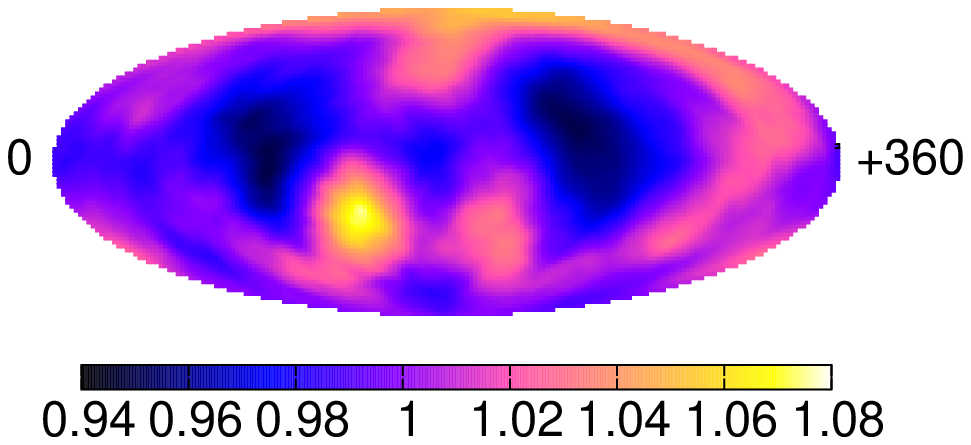}
              \hfil
              \includegraphics[width=0.235\textwidth]{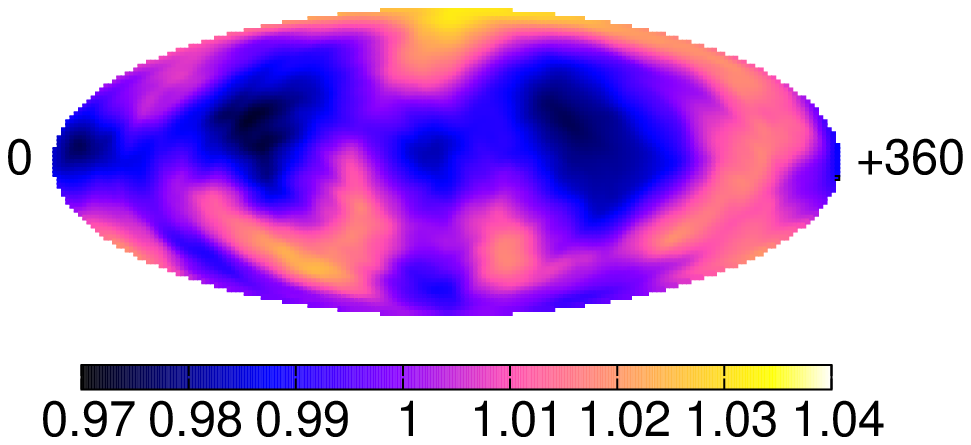}
              \hfil
              \includegraphics[width=0.235\textwidth]{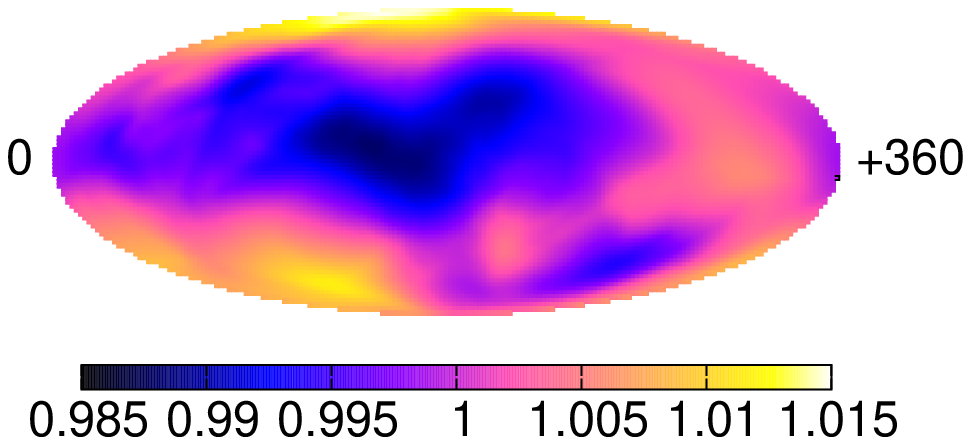}
             }
  \centerline{\includegraphics[width=0.235\textwidth]{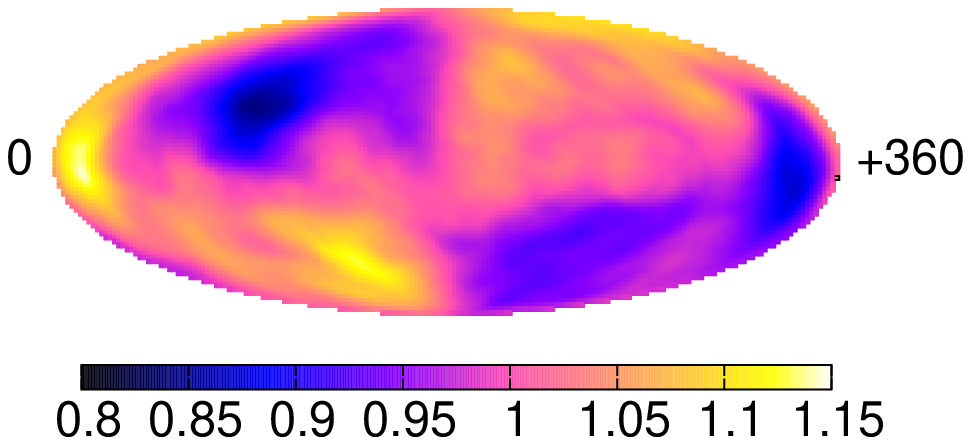}
              \hfil
              \includegraphics[width=0.235\textwidth]{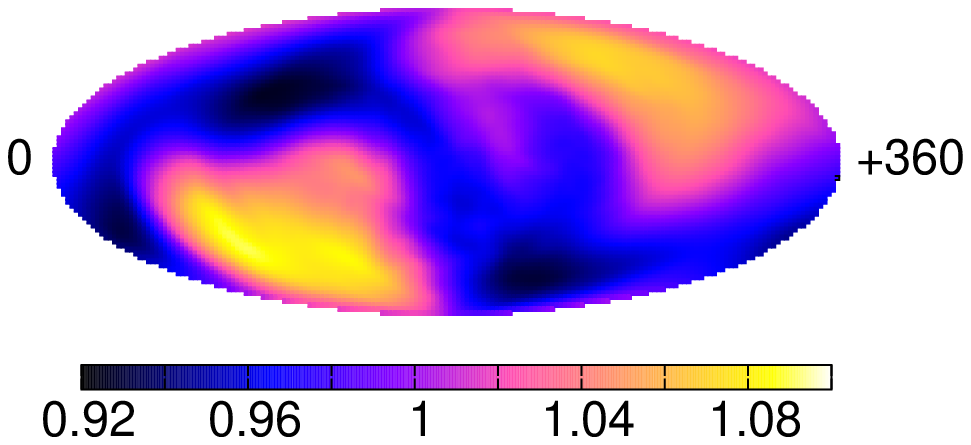}
              \hfil
              \includegraphics[width=0.235\textwidth]{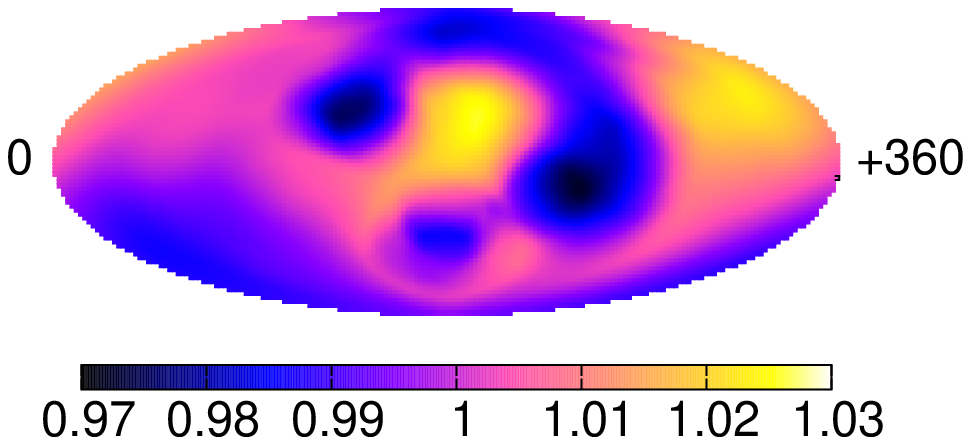}
              \hfil
              \includegraphics[width=0.235\textwidth]{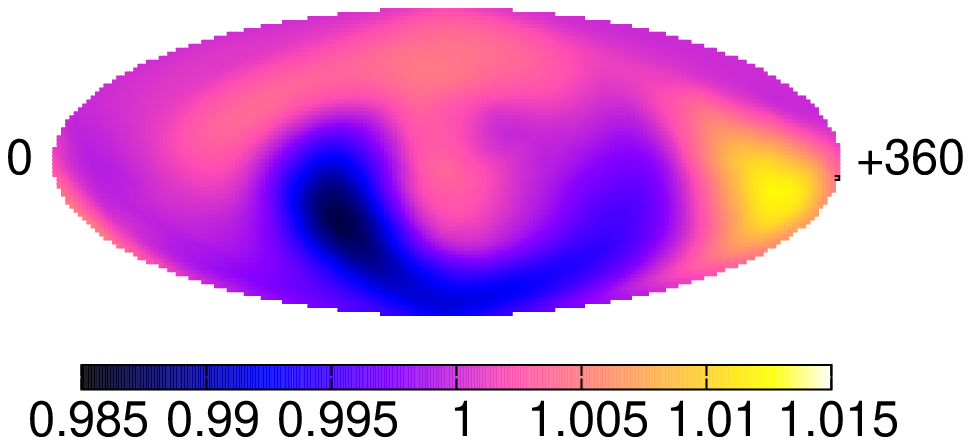}
             }
  \caption{Same as Fig.~\ref{AllSkySmallScales}, for boundary conditions imposed on concentric spheres around Earth with radii $R=100,\,50,\,25,\,10\,$pc (\textit{resp. first, second, third and fourth columns}). \textit{Upper row:} $p/Z=10^{16}$\,eV; \textit{Lower row:} $p/Z=5 \times 10^{16}$\,eV.}
\label{DepdceRad}
\end{figure*}

Fig.~\ref{DepdceRad} shows how the predicted anisotropies in Fig.~\ref{AllSkySmallScales} depend on the maximal back-tracking distance $R$. At 10\,PeV and 50\,PeV, convergence of the sky maps is essentially achieved for $R\gtrsim25$\,pc and $R\gtrsim50$\,pc, respectively. These length scales roughly correspond to the values of $\lambda(p)$ in the vicinity of Earth in the given magnetic field realization. Averaged over many realizations, for $\delta=1/3$, their ratio, as well as the ratio of corresponding anisotropy amplitudes in Fig.~\ref{AllSkySmallScales}, should equal $5^{1/3}$ according to Eq.~(\ref{eq:diff_coeff}), and Eqs.~(\ref{eq:dipole_num}) and~(\ref{eq:a_lm}), respectively. The small deviation from this average of the ratios simulated in the given realization is thus due to ``cosmic variance''. Fig.~\ref{DepdceRad} shows that the small scale fluctuations arise from the local field within $\sim \lambda(p)$, as predicted above.

\begin{figure*}[!t]
  \centerline{\includegraphics[width=0.235\textwidth]{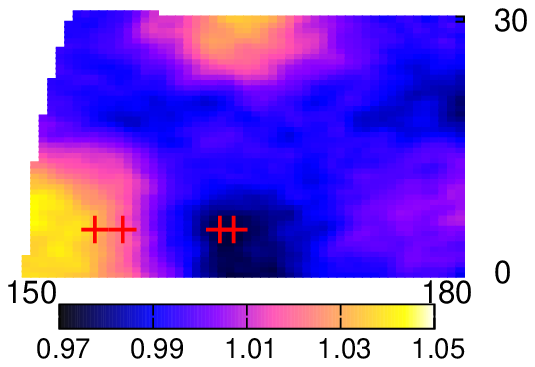}
              \hfil
	      \includegraphics[width=0.235\textwidth]{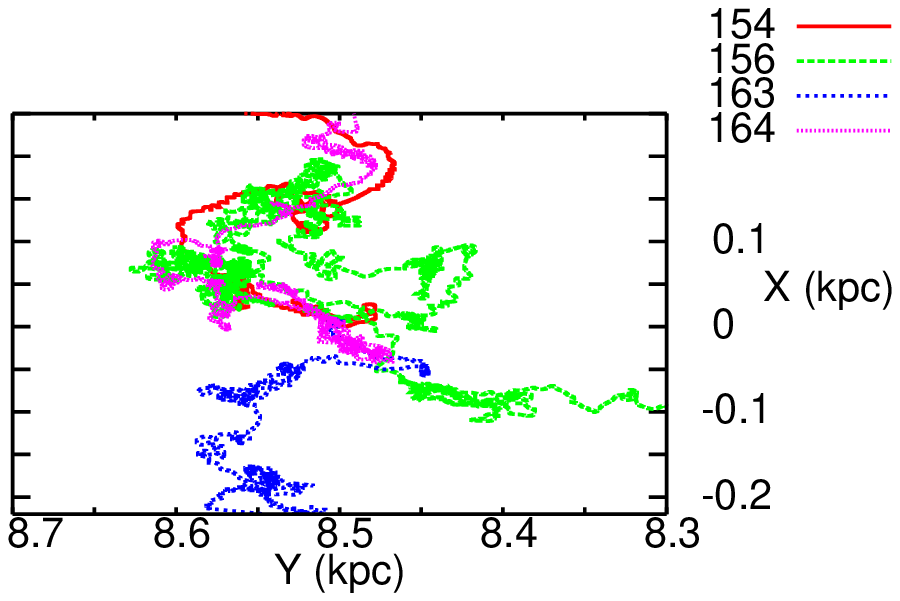}
              \hfil
              \includegraphics[width=0.235\textwidth]{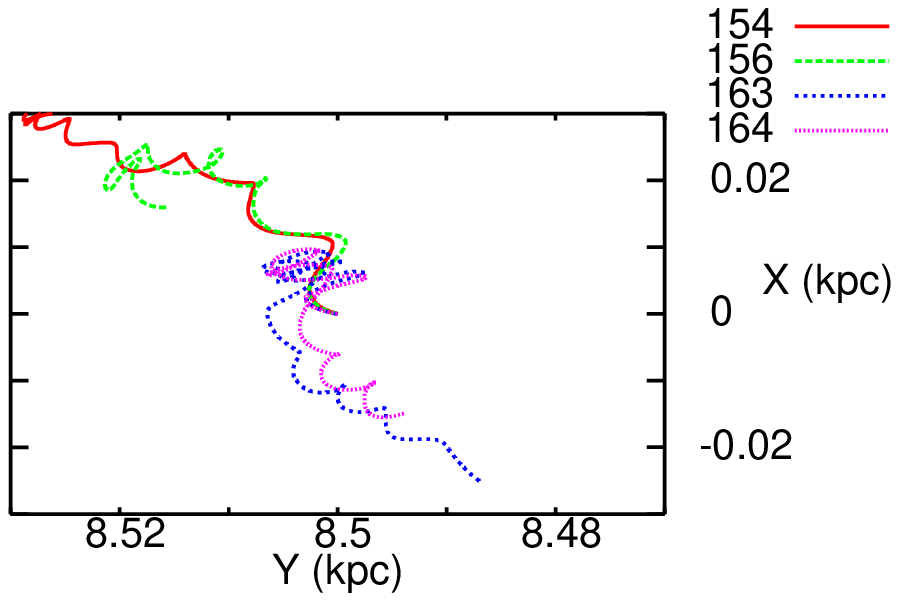}
              \hfil
              \includegraphics[width=0.235\textwidth]{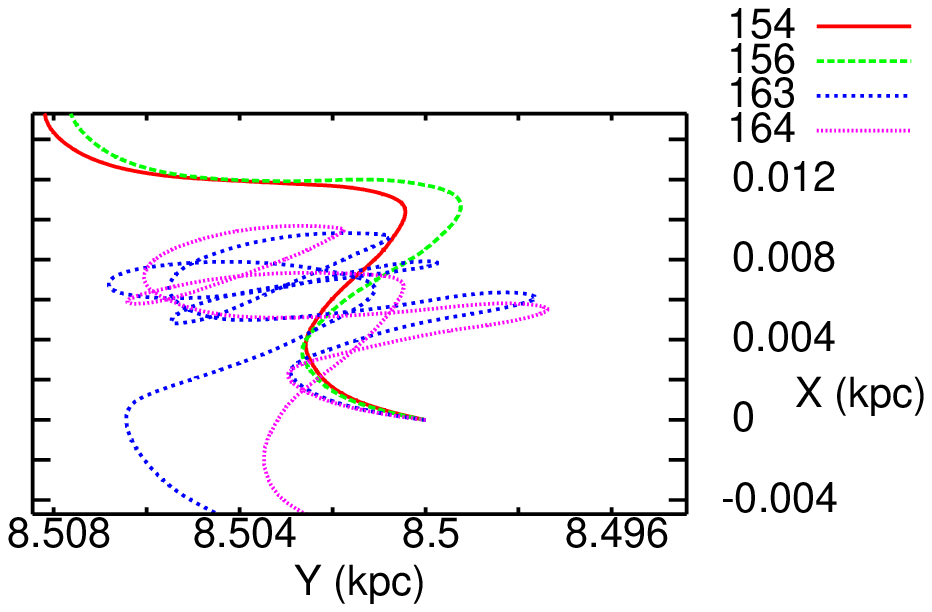}
             }
  \caption{\textit{First panel:} Renormalized CR flux predicted at Earth for $p/Z=10^{16}\,$eV in the sky patch $l=150^{\circ}-180^{\circ}$ and $b=0^{\circ}-30^{\circ}$, smoothed on $5^{\circ}$ radius circles. This is a blow-up of Fig.~\ref{AllSkySmallScales} (left panel); \textit{Second, third and fourth panels:} Trajectories of four CRs back-tracked from Earth, projected onto the $XY$ plane and within $\simeq 200,\,\simeq 20$, and a few pc from Earth (\textit{resp. second, third and fourth panels}). The CR initial directions, denoted by red crosses on the first panel, are $l=154^{\circ},156^{\circ},163^{\circ},164^{\circ}$ (see values in panel labels) and $b=5^{\circ}$ in celestial coordinates.}
\label{3D_Traj}
\end{figure*}

Fig.~\ref{3D_Traj} (first panel) presents CR flux anisotropies in a $30^{\circ} \times 30^{\circ}$ sky patch, after smoothing on $5^{\circ}$ radii circles. The three other panels show the trajectories of four CRs arriving at the red crosses in the first panel (two chosen in an excess region and two in a deficit region). The third and fourth panels show that, at distances $R\lesssim\lambda(p)$ from Earth, the two trajectories arriving in the hot spot tend to come from the direction of the CR density gradient ($Y>8.5$\,kpc) while the other two come from the opposite direction, consistent with Eq.~(\ref{eq:rho}). On larger scales (second panel), the initial directions are more uniformly distributed, again consistent with Eq.~(\ref{eq:rho}): The small scales reflect the last part of the particle trajectories ($\lesssim \lambda(p)/c_0$), before they are detected by the point-like observer.

We verified that small scale anisotropies also appear for anisotropic CR scattering, by performing simulations with an additional large scale field.

\textit{Conclusions.}---We have shown that the observed intermediate and small scale anisotropies in the Galactic CR arrival directions can be naturally explained as the consequence of CR propagation in a turbulent magnetic field. The observed anisotropies could thus be one of the first direct manifestations of the turbulent Galactic magnetic field within the scattering lengths of TeV--PeV CRs, and thus within a few tens of parsecs from Earth. Formally, this effect could have similarities with the ``CR scintillations'' in the inner heliosphere~\cite{Jokipii}. In the future, this should allow new insights into the CR transport in our Galaxy and contribute to our knowledge of the structure of local --and notably interstellar-- magnetic fields.

The authors thank P.~Blasi, J.~R.~Jokipii, M.~Kachelrie\ss, M.~Pohl and D.~Semikoz for useful discussions. This work was supported by the Deutsche Forschungsgemeinschaft through the collaborative research centre SFB 676. GG acknowledges support from the Research Council of Norway, through the Yggdrasil Grant No. 211154/F11. GS acknowledges support from the State of Hamburg, through the Collaborative Research program ``Connecting Particles with the Cosmos'' and from the ``Helmholtz Alliance for Astroparticle Phyics HAP'' funded by the Initiative and Networking Fund of the Helmholtz Association.

\end{document}